\documentclass[pra,preprint]{revtex4}
\usepackage[dvips]{graphicx}

\newcommand{\cm}{cm$^{-1}$}
\newcommand{\tref}[1]{Table~\ref{#1}}
\newcommand{\fref}[1]{Figure~\ref{#1}}

\begin{document}

\title{The Generalized Relativistic Effective Core Potential Calculations 
of the Adiabatic Potential Curve and Spectroscopic Constants 
for the Ground Electronic State of the Ca$_2$}

\author{N.\ S.\ MOSYAGIN}\email{mosyagin@pnpi.spb.ru}
                       \homepage{http://www.qchem.pnpi.spb.ru}
\author{A.\ N.\ PETROV}
\altaffiliation [Also at ] {Division of Quantum Mechanics, St.Petersburg State University, 198904, Russia}
\author{A.\ V.\ TITOV}
\author{A.\ V.\ ZAITSEVSKII}
\altaffiliation [Also at ] {NRC ``Kurchatov Institute'', 1 Kurchatov sq., Moscow, 123182 Russia}

\affiliation{Petersburg Nuclear Physics Institute, 
             Gatchina, St.-Petersburg district 188300, Russia}

\date{\today}

\begin{abstract}
The potential curve, dissociation energy, equilibrium internuclear 
distance, and spectroscopic constants for the $^1\Sigma_g^+$ 
ground state of the Ca$_2$ molecule are calculated with the help of 
the generalized relativistic effective core potential method 
which allows one to exclude the inner core electrons from the calculations 
and to take the relativistic effects into account effectively. 
Extensive generalized correlation basis sets were constructed and employed. 
The scalar relativistic coupled cluster method with corrections for 
high-order cluster amplitudes is used for the correlation treatment. 
The obtained results are analyzed and compared with 
the experimental data and corresponding all-electron results. 

\vspace{2cm}
{\bf SHORT NAME:} GRECP calculations of Ca$_2$


\end{abstract}

\maketitle

\section{Introduction.}

Recently we have reported calculations of the potential curve, 
dissociation energy, equilibrium internuclear distance, and spectroscopic 
constants for the ground state of the Yb$_2$ molecule \cite{Mosyagin:11}.
To estimate reliably the computational accuracy,
one has to analyze carefully the errors of all the used approximations. 
A direct way to calculate
these errors is to carry out the calculations both with and without 
a given approximation. 
As a rule, this way is impractical because the approximations 
are normally introduced just to make the computations feasible.
An alternative way is to compare the calculated results with 
the corresponding experimental data. 
However, the experimental data are not always accessible
or have relatively large errors as is in the case of Yb$_2$.
One can then perform the calculations on the well-studied system with 
similar electronic structure within the same approximations
having in mind a reasonable suggestion that the errors for similar systems
will be close to each other.
As an appropriate analog of Yb$_2$, we have chosen the Ca$_2$ molecule 
that was studied in details
both experimentally \cite{Balfour:75, Vidal:80, Bondybey:84, Allard:02}
and theoretically \cite{Pacchioni:82, Dyall:92a, Porsev:02, Iron:03, 
Czuchaj:03a, Bussery:03, Porsev:06, Patkowski:07, Bouissou:10, Heaven:11, 
Yang:12, Mahapatra:12}.

Similar ground states with the closed-shell ($ns^2$) valence configuration and 
similar first excited $ns^1 np^1 (^3P_0,^3P_1,^3P_2)$ terms
are observed for the Ca ($n=4$) and Yb ($n=6$) atoms \cite{Moore:58}.
These atoms also have similar $(n-1)s^2 (n-1)p^6$ outercore 
and lowest-lying $(n-1)d^0$ virtual shells.
The main differences are the presence of the closed $4f^{14}$ outercore 
(or subvalence) shell and notably stronger relativistic effects in Yb. 
However, the contribution from ``unfreezing'' the $4f$ shell to the binding 
energy is rather small (about 3\%). 
This can be easily understood because the $4f$ shell has a significantly 
smaller radius and essentially lower orbital energy than those for $6s$ 
(and $6p$) ones.

The experimental 
investigations \cite{Balfour:75, Vidal:80, Bondybey:84, Allard:02, Yang:09}
of Ca$_2$ were motivated by considerable interest in 
understanding the van-der-Waals interactions, 
interpretation of stellar absorption spectra, 
studying the metal vapor excimer laser systems,
prediction on cold collision phenomena, etc.
Several fields in ultracold atom physics 
such as photoassociation \cite{Zinner:00,Vogt:07}, 
optical frequency standards \cite{Wilpers:02,Degenhardt:05} 
and possible Bose-Einstein condensates \cite{Grunert:02,Hansen:06}
have been initiated during the last decade
with magneto-optical traps \cite{Zinner:00,Binnewies:01}. 
Interatomic interaction potentials are required for these investigations.
The theoretical studies are additionally aimed to test the modern
quantum chemical methods for this computationally difficult system
(multiconfigurational nature of the ground state, 
weak van-der-Waals interaction, 
noticeable relativistic corrections). 
Some of the above cited authors \cite{Patkowski:07, Yang:12} 
carried out all-electron (AE) molecular calculations with the help of 
the same correlation method as in the present paper that allows us to 
estimate
the accuracy of the more economical generalized relativistic effective core 
potential (GRECP) method in reproducing the all-electron molecular results.

This paper presents the results of our calculations on
the dissociation energy, equilibrium internuclear distance and 
spectroscopic constants for the Ca$_2$ molecule using the GRECP method,
extremely flexible generalized correlation basis sets, correlation treatment 
by the coupled cluster method with single, double and non-iterative-triple 
cluster amplitudes (CCSD(T)) and contributions from higher-order cluster 
amplitudes in comparison with the experimental data and 
corresponding all-electron results.

 \section{Calculations and discussion.}

Scalar relativistic calculations were performed within the GRECP 
model \cite{Titov:99, Petrov:04b, Mosyagin:05a, Mosyagin:05b} 
using the CCSD(T) method (implemented in the 
{\sc molcas} program package \cite{MOLCAS}) for correlation treatment. 
The $4s$ shell of the Ca atom is usually considered as the valence one. 
The $3s$ and $3p$ shells were considered as the outercore ones in 
the GRECP generation procedure \cite{Titov:99}. 
Thus, we use the GRECP with 10 explicitly treated electrons for each Ca atom.
In a series of preliminary calculations, we have estimated the contributions 
from correlations with different shells of Ca to the dissociation energy 
of Ca$_2$ (some of them are presented in \tref{Ca_2}).
The main contribution is provided by the $4s$ shell whereas the contributions 
from the $3s$ and $3p$ shells are relatively small. 
This enabled us to ``freeze'' $3s,3p$ shells in the
starting set of calculations marked as ``4e'' and to incorporate other 
correlation effects through appropriate corrections. 
It is clear that the corresponding contributions from the innermore 
$2p$, $2s$, and $1s$ shells will be significantly smaller so that 
their complete exclusion from the correlation treatment within the 
GRECP model seems to be well justified. 
Generalized correlation basis sets comprising
$(22,22,21,6,4,4,1)/[10,8,9,6,4,4,1]$ functions, basis set C (core),
in the former and $(12,13,9,6,3,3,2)/[12,13,9,6,3,3,2]$, basis set L (large),
$(12,13,9,6,3)/[4,4,4,1,1]$, basis set M (medium, with the $h$
and $i$ harmonics removed from the previous uncontracted basis set),
in the latter cases were constructed by the procedure developed 
previously \cite{Mosyagin:00, Mosyagin:01b}.
For the basis set C construction, the large number of the states of 
the Ca atom and its cation, the leading configurations of which differ by 
the occupation numbers of the $3d$, $4s$, $4p$ valence spinors and 
have the completely occupied $3s$, $3p$ outer core spinors, were considered.
The basis set construction procedure is designed to account primarily for 
correlations which have different contributions to the states under 
consideration, so that possible omissions in the resulting basis set
cause nearly state-independent errors and give accurate transition energies. 
Therefore, the resulting basis set may be nearly complete in the valence 
region but relatively poor in the outer core region. 
The basis set superposition errors (BSSE) will have weak dependence on
the valence shell configuration, and may be estimated quite 
accurately by the counterpoise correction (CPC) method because
there are no ambiguities in the occupation numbers for the outer core shells
in the effective state of ``atom-in-molecule''. 

Calculations were carried out for the internuclear distances ($R$)
presented in \tref{tabca2} (from 6 to 13~a.u.\ and for 100~a.u.). 
All our results were rectified using the 
CPC \cite{Gutowski:86, Liu:89} calculated for the Ca $4s^2$ state
with one more Ca atom treated as a ghost one. 
The stage of calculation of the molecular constants \cite{Mitin:98} 
begins with fitting the numerical potential curve for the dimer
by polynomials with the help of the quasi-Hermitian method. 
Appropriate derivatives of the potential curve at the equilibrium point 
are calculated by recurrence relations.
Then rovibrational Schr{\"o}dinger equation is solved by the Dunham method 
to express the Dunham coefficients in terms of these derivatives.

\begin{figure}[ht]
   \includegraphics[width=0.9\textwidth]{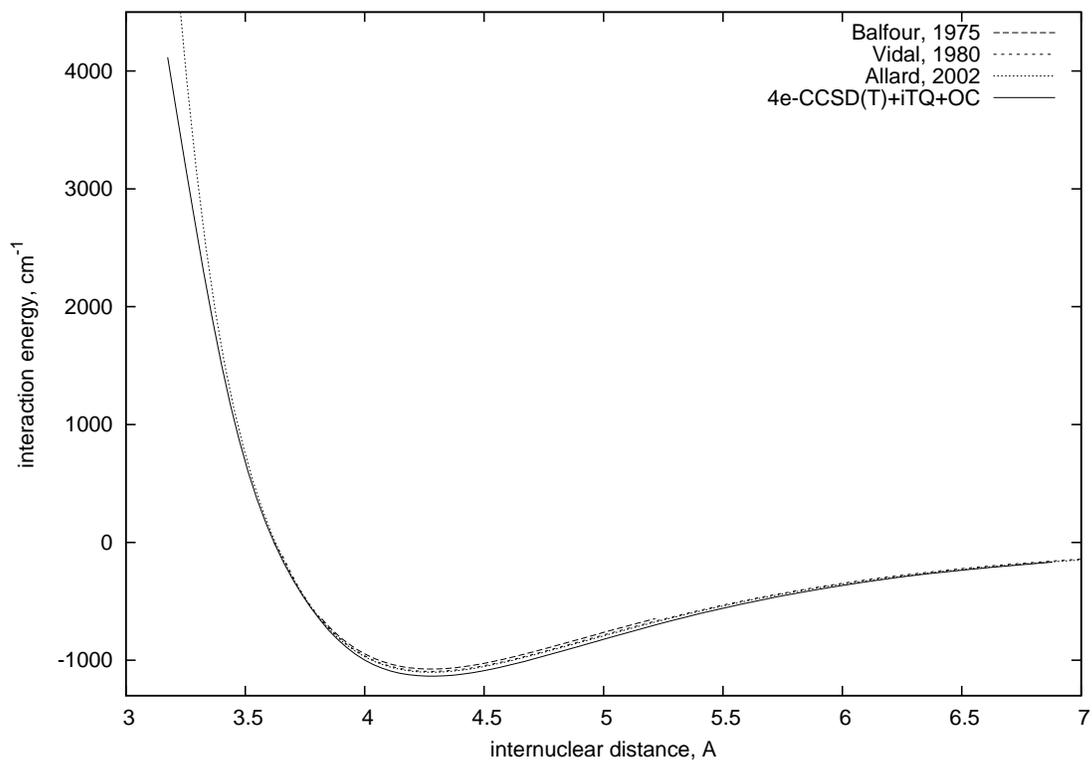}
\caption{ \label{figca2} 
The calculated (4e-CCSD(T)+iTQ+OC) and experimentally derived 
(Balfour, 1975 \cite{Balfour:75}; Vidal, 1980 \cite{Vidal:80}; 
Allard, 2002 \cite{Allard:02}) potential energy functions for 
the Ca$_2$ ground state. 
}
\end{figure}

The $^1\Sigma_g^+$ closed-shell ground state of the Ca$_2$ molecule
disscociates into two Ca atoms in the $4s^2 (^1S)$ ground state.
The computed ground-state potential energy curves 
for the Ca$_2$ molecule are shown in \tref{tabca2} and \fref{figca2}.
One can see from \fref{figca2} that our final potential curve
is in a good agreement with the experimentally derived curves in papers 
\cite{Balfour:75, Vidal:80, Allard:02}.
Our results for the dissociation energy, equilibrium internuclear
distance, and main spectroscopic constants are listed in \tref{Ca_2}.
Following the scheme of our previous Yb$_2$ calculation \cite{Mosyagin:11},
we started from 4-electron scalar relativistic CCSD(T) (denoted as 4e-CCSD(T) 
below) calculations with rather large basis set L, which gave $D_e=$1022~\cm\ 
(that is about 1.5 times larger than that for Yb$_2$). 
The negligible CPC (0.3~\cm\ for 
dissociation energy) indicates a good quality of the basis set used. 
Subsequent calculations of the effects of the difference between 
the iterative and non-iterative triple cluster amplitudes 
(CCSDT-CCSD(T) or contribution from iteration of triples) 
as well as of quadruple cluster amplitudes
(these two contributions are denoted further as the iTQ contribution), and
valence -- outer core correlations (OC)
described below have shown that the corresponding contributions to the Ca--Ca
interaction energy are within 17\% 
(with respect to our final dissociation energy of 1136~\cm),
thus, justifying the choice of the 4e-CCSD(T)
scheme as a good initial approximation. Note that the 4-electron FCI or
20-electron CCSD(T) calculations with considerably smaller 
basis sets M or C have given essentially lower $D_e$ 
estimates ($D_e=$959 or 930~\cm, correspondingly).
Thus, the quality of the basis set is important for accurate calculations 
of the calcium dimer.

The contribution from the quadruple cluster amplitudes 
together with the difference between the iterative and 
non-iterative triple amplitudes, $\Delta E_{iTQ}$, 
was estimated as the difference between 
the energy lowerings $E={\cal E}(Ca_2)-2{\cal E}(Ca)$
obtained in the 4e-FCI and 4e-CCSD(T) 
calculations with basis set M 
for each of the above mentioned internuclear distances
\[
\Delta E^M_{4e-iTQ}(R) = E^M_{4e-FCI}(R) - E^M_{4e-CCSD(T)}(R),
\]
where ${\cal E}(Ca_2)$ and ${\cal E}(Ca)$ are the total energies 
calculated in the molecular basis set for the Ca$_2$ molecule and 
the Ca atom, respectively. 
Thus, this contribution takes into account the CPC.
This difference was then added to the total energy
obtained in the 4e-CCSD(T) calculation with basis set L
\[
E_{4e-CCSD(T)+iTQ}(R)=E^L_{4e-CCSD(T)}(R)+\Delta E^M_{4e-iTQ}(R).
\]
The dissociation energy, equilibrium internuclear distance, and spectroscopic 
constants were calculated with the obtained 4e-CCSD(T)+iTQ potential curve.
The derived correction from the iTQ amplitudes 
to the dissociation energy, 190~\cm, is 17\% 
with respect to our final $D_e$ value
(that is also about 1.5 times larger by absolute value than that for Yb$_2$ 
but has the same relative value).

The contribution from the correlations with the $3s$ and $3p$ 
electrons, $\Delta E_{OC}$,
was estimated as the difference between the energy lowerings
found in the 20e-CCSD(T) and 4e-CCSD(T) calculations 
with basis set C for each of the above mentioned internuclear distances
\[
\Delta E^{C,CCSD(T)}_{OC}(R) = E^C_{20e-CCSD(T)}(R) - E^C_{4e-CCSD(T)}(R).
\]

The only difference between these two calculations is the number of 
correlated electrons, therefore, the differences in the energy lowerings
give the contribution of the OC correlations.
These differences were then added to the 4e-CCSD(T) and 4e-CCSD(T)+iTQ 
energy lowerings derived above: 
\[
E_{4e-CCSD(T)+OC}(R)=E^L_{4e-CCSD(T)}(R)+\Delta E^{C,CCSD(T)}_{OC}(R),
\]
\[
E_{4e-CCSD(T)+iTQ+OC}(R)=E_{4e-CCSD(T)+iTQ}(R)+\Delta E^{C,CCSD(T)}_{OC}(R).
\]
The 4e-CCSD(T)+OC and 4e-CCSD(T)+iTQ+OC dissociation energy, equilibrium 
internuclear distance and 
spectroscopic constants were calculated with the obtained potential curves.
The dissociation energy was decreased by 76~\cm\ 
(that is about 1.4 times larger than that for Yb$_2$), 
whereas the corresponding CPC contribution was obtained as about 10~\cm. 

It should be noted that the contribution from the spin-dependent 
interactions for the excluded innercore 2p shell of Ca is effectively taken 
into account by the GRECP method.
The contribution from the spin-dependent interactions for 
the outercore, valence and virtual shells of Ca is neglected 
in the present scalar relativistic calculations.
We estimated this contribution as the difference between the energy lowerings 
in the 20e-CCSD calculations with the full spin-dependent and 
spin-averaged GRECP operators \cite{Titov:99} 
with the $(22,22,21,6)/[5,4,3,2]$ basis set
for the internuclear distance R=8.0~a.u.\ 
(that is close to the equilibrium one). 
The CPC was also taken into account.
The contribution was significantly less than one wave number.

It should be noted that the low convergence threshold of $10^{-8}$ and 
the approximation of the potential curve by analytic polynomials were used.
The errors in the calculated total energies have rather systematic nature
(due to neglecting the unaccounted effects) than a random one, therefore, 
the large number of points is not necessary for good statistics.
To check the saturation in the number of the potential curve points,
we have repeated the dissociation energy, equilibrium internuclear
distance and spectroscopic constant calculations without two extreme points 
(corresponding to 6 and 13~a.u.). 
As one can see from \tref{Ca_2}, the given data are almost unchanged.
Similar situation was observed for different analytical functions 
(Legendre, second kind Chebyshev and power polynomials) used for 
interpolation in program \cite{Mitin:98} instead of the Laguerre polynomials. 
The latters provided the best approximation in the least square sense and 
were used for the calculation of the data in \tref{Ca_2}.

If the same correlation method and basis sets of similar flexibility are used,
one can compare the GRECP and all-electron results.
The nonrelativistic AE/CCSD(T) dissociation energies extrapolated 
to the complete basis set limit and with accounting for the CPC were 
calculated in \cite{Patkowski:07, Yang:12} as 993 and 997 \cm.
The relativistic effects are taken into account in the GRECP method 
by construction.
Moreover, interplay of the relativistic and correlation effects is also 
taken into account in the correlation GRECP calculations.
The relativistic effects were added as corrections 
in \cite{Patkowski:07, Yang:12}. 
They decreased the dissociation energy on 37 \cm\ in the both above studies.
Thus, our 4e-CCSD(T)+OC dissociation energy of 939 \cm\ is in a good 
agreement with the corrected 956 and 960 values from the above 
all-electron calculations.
It should be noted that this GRECP result was obtained with rather large 
but a finite basis set without the complete basis set limit extrapolation.
The all-electron dissociation energies calculated 
in \cite{Patkowski:07, Yang:12} with the largest basis sets and CPCs 
(but without the extrapolation to the complete basis set limit) 
are by 10 and 41 \cm\ lower.
We conclude that the small difference (about 20 \cm) in the GRECP and 
all-electron results can be mainly due to both the GRECP errors and 
incompleteness of our basis set. 
The approximate accounting for the relativistic effects 
in \cite{Patkowski:07, Yang:12} can also contribute to this difference.
This difference is significantly smaller than the unaccounted iTQ 
contribution for the OC shells.
We estimated the iT contribution for the OC shells with the help of 
the 20e-CCSDT (i.e.\ with the iterative triple amplitudes), 20e-CCSD(T) 
4e-CCSDT, and 4e-CCSD(T) calculations with the $(22,22,21,6)/[5,4,3,2]$ 
basis set for the internuclear distance R=8.0~a.u.\ as
\[
\Delta E_{iT\&OC}(R) = E_{20e-CCSDT}(R) - E_{20e-CCSD(T)}(R) 
                     - E_{4e-CCSDT}(R) +  E_{4e-CCSD(T)}(R),
\]
The CPC was also taken into account.
This contribution decreased the dissociation energy by -101~\cm. 
These dissociation energies differ from the final ones by 
the iTQ contribution which were calculated as 197 and 196 \cm\ in 
the present paper and in \cite{Patkowski:07} only for the valence shells
and as 136 \cm\ in \cite{Yang:12} for both the valence and outercore shells
but without the CPC.

\section{Conclusions.}

One can see that the GRECP method allows one to reproduce perfectly
the corresponding all-electron results from \cite{Patkowski:07, Yang:12}.
Our final results are presented in the 4e-CCSD(T)+iTQ+OC line in \tref{Ca_2}.
The very small differences between these results and the experimental data 
are mainly due to neglecting the iTQ contribution for the OC shells
which will decrease the dissociation energy. The errors due to the 
incompleteness of our basis sets and the errors of the GRECP method 
are expected to be essentially smaller.
A good agreement of our results with the experimental data should not be 
considered as fortuitous coincidence because it is observed not only 
for one parameter (such as $D_e$) but also for several independent 
parameters ($R_e$, $D_e$, $w_e$, $w_e x_e$, $\alpha_e$, $-Y_{02}$).

\begin{acknowledgments} 
The reported study was partially supported by RFBR, research project 
No.\ 10--03--00727a and by Russian Ministry of Education and Science, 
contract No.\ 07.514.11.4141 (2012--2013). 
\end{acknowledgments}

\bibliographystyle{bib/bst/PTCPmax}

\bibliography{bib/JournAbbr,bib/Titov,bib/TitovLib,bib/Kaldor,bib/Isaev,bib/TitovAbs,bib/MosyaginLib}

\newcommand{\noopsort}[1]{} \newcommand{\printfirst}[2]{#1}
  \newcommand{\singleletter}[1]{#1} \newcommand{\switchargs}[2]{#2#1}
\begin{thebibliography}{10}
\expandafter\ifx\csname url\endcsname\relax
  \def\url#1{{\tt #1}}\fi
\expandafter\ifx\csname urlprefix\endcsname\relax\def\urlprefix{URL }\fi
\providecommand{\eprint}[2][]{\url{#2}}

\bibitem{Mosyagin:11}
Mosyagin, N.~S., Petrov, A.~N., \protect\BIBand{} Titov, A.~V.
\newblock The effect of the iterative triple and quadruple cluster amplitudes
  on the adiabatic potential curve in the coupled cluster calculations of the
  ground electronic state of the {Yb} dimer.
\newblock {\em Int.\ J.\ Quantum Chem.\/} {\bf 111}, 3793--3798 (2011).

\bibitem{Balfour:75}
Balfour, W.~J. \protect\BIBand{} Whitlock, R.~F.
\newblock The visible absorption spectrum of diatomic calcium.
\newblock {\em Can.\ J.\ Phys.\/} {\bf 53}, 472--485 (1975).

\bibitem{Vidal:80}
Vidal, C.~R.
\newblock The molecular constants and potential energy curves of the {Ca}$_2$
  {$A^1\Sigma^+_u$}--{$X^1\Sigma^+_g$}.
\newblock {\em J.\ Chem.\ Phys.\/} {\bf 72}, 1864--1874 (1980).

\bibitem{Bondybey:84}
Bondybey, V.~E. \protect\BIBand{} English, J.~H.
\newblock Laser-induced fluorescence of the calcium dimer in supersonic jet:
  the red spectrum of {Ca}$_2$.
\newblock {\em Chem.\ Phys.\ Lett.\/} {\bf 111}, 195--200 (1984).

\bibitem{Allard:02}
Allard, O., Pashov, A., {Kn\"ockel}, H., \protect\BIBand{} Tiemann, E.
\newblock Ground-state potential of the {Ca} dimer from fourier transform
  spectroscopy.
\newblock {\em Phys. Rev. A\/} {\bf 66}, 042503 (2002).

\bibitem{Pacchioni:82}
Pacchioni, G. \protect\BIBand{} Koutecky, J.
\newblock The bond nature of alkaline-earth homonuclear metal clusters
  investigated with pseudopotential {CI} method.
\newblock {\em Chem.\ Phys.\/} {\bf 71}, 181--198 (1982).

\bibitem{Dyall:92a}
Dyall, K.~G. \protect\BIBand{} McLean, A.~D.
\newblock The potential energy curves of the {$X~^1\Sigma^+_g$} ground states
  of {Mg}$_2$ and {Ca}$_2$ using the interacting correlated fragments model.
\newblock {\em J.\ Chem.\ Phys.\/} {\bf 97}, 8424--8431 (1992).

\bibitem{Porsev:02}
Porsev, S.~G. \protect\BIBand{} Derevianko, A.
\newblock High-accuracy relativistic many-body calculations of van der {W}aals
  coefficients {C}$_6$ for alkaline-earth-metal atoms.
\newblock {\em Phys. Rev. A\/} {\bf 65}, 020701(R) (2002).

\bibitem{Iron:03}
Iron, M.~A., Oren, M., \protect\BIBand{} Martin, J. M.~L.
\newblock Alkali and alkaline earth metal compounds: core-valence basis sets
  and importance of subvalence correlation.
\newblock {\em Mol.\ Phys.\/} {\bf 101}, 1345--1361 (2003).

\bibitem{Czuchaj:03a}
Czuchaj, E., {Kr\'osnicki}, M., \protect\BIBand{} Stoll, H.
\newblock Valence ab initio calculation of the potential-energy curves for the
  {Ca}$_2$ dimer.
\newblock {\em Theor.\ Chem.\ Acc.\/} {\bf 110}, 28--33 (2003).

\bibitem{Bussery:03}
Bussery-Honvault, B., Launay, J.-M., \protect\BIBand{} Moszynski, R.
\newblock Cold collisions of ground-state calcium atoms in a laser field: A
  theoretical study.
\newblock {\em Phys. Rev. A\/} {\bf 68}, 032718 (2003).

\bibitem{Porsev:06}
Porsev, S.~G. \protect\BIBand{} Derevianko, A.
\newblock High-accuracy calculations of dipole, quadrupole, and octupole
  electric dynamic polarizabilities and van der {W}aals coefficients {C}$_6$,
  {C}$_8$, and {C}$_{10}$ for alkaline-earth dimers.
\newblock {\em Sov. Phys. JETP\/} {\bf 102}, 195--205 (2006).

\bibitem{Patkowski:07}
Patkowski, K., Podeszwa, R., \protect\BIBand{} Szalewicz, K.
\newblock Interactions in diatomic dimers involving closed-shell metals.
\newblock {\em J.\ Phys.\ Chem.\ A\/} {\bf 111}, 12822--12838 (2007).

\bibitem{Bouissou:10}
Bouissou, T., Durand, G., Heitz, M.-C., \protect\BIBand{} Spiegelman, F.
\newblock A comprehensive theoretical investigation of the electronic states of
  {Ca}$_2$ up to the {Ca$(4s^2~^1S)$}$+${Ca$(4s5p~^1P)$} dissociation limit.
\newblock {\em J.\ Chem.\ Phys.\/} {\bf 133}, 164317 (2010).

\bibitem{Heaven:11}
Heaven, M.~C., Bondybey, V.~E., Merritt, J.~M., \protect\BIBand{} Kaledin,
  A.~L.
\newblock The unique bonding characteristics of beryllium and the group {IIA}
  metals.
\newblock {\em Chem.\ Phys.\ Lett.\/} {\bf 506}, 1--14 (2011).

\bibitem{Yang:12}
Yang, D.-D. \protect\BIBand{} Wang, F.
\newblock Theoretical investigation for spectroscopic constants of ground-state
  alkaline-earth dimers with high accuracy.
\newblock {\em Theor.\ Chem.\ Acc.\/} {\bf 131}, 1117 (2012).

\bibitem{Mahapatra:12}
Mahapatra, U.~S. \protect\BIBand{} Chattopadhyay, S.
\newblock Single reference coupled cluster calculations for weakly bound
  alkaline-earth metal dimers in the ground state: a useful perturbative scheme
  for an iterative triples correction.
\newblock {\em Mol.\ Phys.\/} {\bf 110}, 75--83 (2012).

\bibitem{Moore:58}
Moore, C.~E.
\newblock {\em Atomic Energy Levels\/}, vol. 1-3 (Natl. Bur. Stand. (US), Circ.
  No. 467, Washington, 1958).

\bibitem{Yang:09}
Yang, D.~D., Li, P., \protect\BIBand{} Tang, K.~T.
\newblock The ground state van der {W}aals potentials of the calcium dimer and
  calcium rare-gas complexes.
\newblock {\em J.\ Chem.\ Phys.\/} {\bf 131}, 154301 (2009).

\bibitem{Zinner:00}
Zinner, G., Binnewies, T., \protect\BIBand{} Riehle, F.
\newblock Photoassociation of cold ca atoms.
\newblock {\em Phys. Rev. Lett.\/} {\bf 85}, 2292--2295 (2000).

\bibitem{Vogt:07}
Vogt, F., Grain, C., Nazarova, T., Sterr, U., Riehle, F., Lisdat, C.,
  \protect\BIBand{} Tiemann, E.
\newblock Determination of the calcium ground state scattering length by
  photoassociation spectroscopy at large detunings.
\newblock {\em Eur. Phys. J. D\/} {\bf 44}, 73--79 (2007).

\bibitem{Wilpers:02}
Wilpers, G., Binnewies, T., Degenhardt, C., Sterr, U., Helmcke, J.,
  \protect\BIBand{} Riehle, F.
\newblock Optical clock with ultracold neutral atom.
\newblock {\em Phys. Rev. Lett.\/} {\bf 89}, 230801 (2002).

\bibitem{Degenhardt:05}
Degenhardt, C., Stoehr, H., Lisdat, C., Wilpers, G., Schnatz, H., Lipphardt,
  B., Nazarova, T., Pottie, P.-E., Sterr, U., Helmcke, J., \protect\BIBand{}
  Riehle, F.
\newblock Calcium optical frequency standard with ultracold atoms: Approaching
  $10^{−15}$ relative uncertainty.
\newblock {\em Phys. Rev. A\/} {\bf 72}, 062111 (2005).

\bibitem{Grunert:02}
{Gr\"unert}, J. \protect\BIBand{} Hemmerich, A.
\newblock Sub-doppler magneto-optical trap for calcium.
\newblock {\em Phys. Rev. A\/} {\bf 65}, 041401(R) (2002).

\bibitem{Hansen:06}
Hansen, D. \protect\BIBand{} Hemmerich, A.
\newblock Observation of multichannel collisions of cold metastable calcium
  atoms.
\newblock {\em Phys. Rev. Lett.\/} {\bf 96}, 073003 (2006).

\bibitem{Binnewies:01}
Binnewies, T., Wilpers, G., Sterr, U., Riehle, F., Helmcke, J.,
  {Mehlst\"aubler}, T.~E., Rasel, E.~M., \protect\BIBand{} Ertmer, W.
\newblock Doppler cooling and trapping on forbidden transitions.
\newblock {\em Phys. Rev. Lett.\/} {\bf 87}, 123002 (2001).

\bibitem{Titov:99}
Titov, A.~V. \protect\BIBand{} Mosyagin, N.~S.
\newblock Generalized relativistic effective core potential: {T}heoretical
  grounds.
\newblock {\em Int.\ J.\ Quantum Chem.\/} {\bf 71}, 359--401 (1999).

\bibitem{Petrov:04b}
Petrov, A.~N., Mosyagin, N.~S., Titov, A.~V., \protect\BIBand{} Tupitsyn, I.~I.
\newblock Accounting for the {B}reit interaction in relativistic effective core
  potential calculations of actinides.
\newblock {\em J.\ Phys.\ B\/} {\bf 37}, 4621--4637 (2004).

\bibitem{Mosyagin:05a}
Mosyagin, N.~S., Petrov, A.~N., Titov, A.~V., \protect\BIBand{} Tupitsyn, I.~I.
\newblock {GRECPs} accounting for {B}reit effects in uranium, plutonium and
  superheavy elements 112, 113, 114.
\newblock In: J.-P. Julien, J.~Maruani, D.~Mayou, S.~Wilson, \protect\BIBand{}
  G.~{Delgado-Barrio} (eds.) {\em Recent Advances in the Theory of Chemical and
  Physical Systems\/}, vol. B~15 of {\em Progr.\ Theor.\ Chem.\ Phys.\/}, pp.
  229--251 (Springer, Dordrecht, The Netherlands, 2006).

\bibitem{Mosyagin:05b}
Mosyagin, N.~S. \protect\BIBand{} Titov, A.~V.
\newblock Accounting for correlations with core electrons by means of the
  generalized {RECP}: Atoms {Hg} and {Pb} and their compounds.
\newblock {\em J.\ Chem.\ Phys.\/} {\bf 122}, 234106 (2005).

\bibitem{MOLCAS}
Andersson, K., Blomberg, M. R.~A., {F\"ulscher}, M.~P., {Karlstr\"om}, G.,
  Lindh, R., Malmqvist, P.-A., {Neogr\'ady}, P., Olsen, J., Roos, B.~O.,
  Sadlej, A.~J., {Sch\"utz}, M., Seijo, L., {Serrano-Andr\'es}, L., Siegbahn,
  P. E.~M., \protect\BIBand{} Widmark, P.-O. (1999).
\newblock Quantum-chemical program package {``{\sc molcas}''}, Version 4.1.

\bibitem{Mosyagin:00}
Mosyagin, N.~S., Eliav, E., Titov, A.~V., \protect\BIBand{} Kaldor, U.
\newblock Comparison of relativistic effective core potential and all-electron
  {D}irac-{C}oulomb calculations of mercury transition energies by the
  relativistic coupled-cluster method.
\newblock {\em J.\ Phys.\ B\/} {\bf 33}, 667--676 (2000).

\bibitem{Mosyagin:01b}
Mosyagin, N.~S., Titov, A.~V., Eliav, E., \protect\BIBand{} Kaldor, U.
\newblock Generalized relativistic effective core potential and relativistic
  coupled cluster calculation of the spectroscopic constants for the {HgH}
  molecule and its cation.
\newblock {\em J.\ Chem.\ Phys.\/} {\bf 115}, 2007--2013 (2001).

\bibitem{Gutowski:86}
Gutowski, M., {van Lenthe}, J.~H., Verbeek, J., {van Duijneveldt}, F.~B.,
  \protect\BIBand{} Cha{\l}asi{\'n}ski, G.
\newblock The basis set superposition error in correlated electronic structure
  calculations.
\newblock {\em Chem.\ Phys.\ Lett.\/} {\bf 124}, 370--375 (1986).

\bibitem{Liu:89}
Liu, B. \protect\BIBand{} McLean, A.~D.
\newblock The interacting correlated fragments model for weak interactions,
  basis set superposition error, and the helium dimer potential.
\newblock {\em J.\ Chem.\ Phys.\/} {\bf 91}, 2348--2359 (1989).

\bibitem{Mitin:98}
Mitin, A.~V.
\newblock Calculation of rovibrational energy levels of diatomic molecules by
  {D}unham method with potential obtained from ab~initio calculations.
\newblock {\em J.\ Comput.\ Chem.\/} {\bf 19}, 94--101 (1998).

\end{thebibliography}

\clearpage
\begin{table}
\caption{\label{tabca2}
Potential energy functions for the Ca$_2$ ground state
calculated with the help of the GRECP and different correlation methods. 
Internuclear distances $R$ and total energy lowerings 
$E={\cal E}(Ca_2)-2{\cal E}(Ca)$ are in a.u.
}
\begin{tabular}{rrrr}
\hline
 $R$   & \multicolumn{3}{c}{$E(R)$ from 4e-CCSD(T)+} \\
\cline{2-4}                                        
       & +OC         & +iTQ        & +OC+iTQ     \\
\hline                                             
   6.0 &  0.02033301 &  0.02378395 &  0.01875840 \\
   6.5 &  0.00663326 &  0.00778011 &  0.00523450 \\
   7.0 & -0.00026847 & -0.00049810 & -0.00151446 \\
   7.5 & -0.00327505 & -0.00421945 & -0.00437114 \\
   8.0 & -0.00421751 & -0.00543463 & -0.00516116 \\
   8.5 & -0.00416952 & -0.00539711 & -0.00496276 \\
   9.0 & -0.00371447 & -0.00482274 & -0.00436746 \\
  10.0 & -0.00258660 & -0.00336334 & -0.00301022 \\
  11.0 & -0.00167843 & -0.00218223 & -0.00194408 \\
  12.0 & -0.00105967 & -0.00137987 & -0.00122368 \\
  13.0 & -0.00066214 & -0.00086506 & -0.00076319 \\
 100.0 &  0.00000000 &  0.00000000 &  0.00000000 \\
\hline
\end{tabular}
\end{table}

\begin{table}
\caption{\label{Ca_2} 
The dissociation energy, equilibrium internuclear distance, and 
spectroscopic constants of the $^1\Sigma_g^+$ ground state
of the $^{40}$Ca$_2$ molecule. $R_e$ is in \AA, $B_e$ in $10^{-2}$~\cm, 
$\alpha_e$ in $10^{-4}$~\cm, $Y_{02}$ in $10^{-8}$~\cm, 
and other values in \cm.
}
\begin{tabular}{lllllllll}
\hline 
 Method                                    & $R_e$ & $D_e$        & $w_e$ & $D_0^0$ & $B_e$ & $w_e x_e$ & $\alpha_e$    & $-Y_{02}$     \\
\hline                                                                                                                                    
\multicolumn{9}{c}{Present GRECP calculations:}                                                                                         \\
\hline                                                                                                                                    
 4e-CCSD(T)                                & 4.394 & 1022         &  62.5 &    ~991 &  4.36 &      1.05 & 6.88          & 8.48          \\
 4e-CCSD(T)+OC                             & 4.337 & ~939         &  60.1 &    ~909 &  4.47 &      1.06 & 7.59          & 9.95          \\
 4e-CCSD(T)+iTQ                            & 4.345 & 1212         &  67.4 &    1178 &  4.46 &      1.01 & 6.37          & 7.79          \\
 4e-CCSD(T)+iTQ+OC                         & 4.283 & 1136         &  65.4 &    1104 &  4.59 &      1.02 & 6.97          & 9.03          \\
 ---''--- (2 less points)$^{\rm a}$        & 4.283 & 1136         &  65.4 &    1104 &  4.59 &      1.03 & 6.98          & 9.03          \\
\hline                                                                                                                                    
\multicolumn{9}{c}{Experimental data:}                                                                                                  \\
\hline                                                                                                                                    
 Balfour, 1975 \cite{Balfour:75}           & 4.277 & 1075$\pm$150 &  64.9 &    1043 &  4.61 &      1.07 & 7.03$\pm$0.03 & 9.52$\pm$0.11 \\
 Vidal, 1980 \cite{Vidal:80}               & 4.279 & 1095         &  65.0 &    1063 &  4.61 &      1.08 & 6.97          & 9.07$\pm$0.11 \\
 Bondybey, 1984 \cite{Bondybey:84}         &       & 1095         &  65.1 &    1063 &  4.61 &           & 6.97          &               \\
 Allard, 2002 \cite{Allard:02}             &       & 1102         &       &    1070 &       &           &               &               \\
\hline                                                                                                                                    
\multicolumn{9}{c}{Previous calculations:}                                                                                              \\
\hline                                                                                                                                    
 2e-PP/4e-FCI \cite{Pacchioni:82}          & 5.1   & ~234         &       &         &       &           &               &               \\
 2e-PP/4e-MRCI+QDPT \cite{Bouissou:10}     & 4.29  & ~943         &  61.7 &    ~912 &       &      0.8  &               &               \\
 AE/20e-CCSD(T) \cite{Iron:03}             & 4.326 & ~991         &  61.5 &    ~961 &       &      1.07 &               &               \\
 10e-PP/20e-CCSD(T) \cite{Czuchaj:03a}     & 4.37  & 1015         &  62.4 &    ~984 &       &           &               &               \\
 AE/CCSD(T) \cite{Heaven:11}               & 4.299 & 1034         &  90.0 &    ~989 &       &           &               &               \\
 AE+DPT2/20e-CCSD(T)+iT+(Q) \cite{Yang:12} & 4.287 & 1095         &  63.8 &    1063 &  4.58 &      1.15 & 7.23          &               \\
 AE+BP/SAPT \cite{Bussery:03}              & 4.3   & 1113         &       &    1082 &       &           &               &               \\
 AE+DK/CCSD(T)+iTQ \cite{Patkowski:07}     &       & 1152$\pm$51  &       &         &       &           &               &               \\
 AE/4e-ICF \cite{Dyall:92a}                & 4.342 & 1240         &  67.9 &    1206 &       &      1.06 &               &               \\
 AE/4e-CCSDT-1a+d \cite{Mahapatra:12}      & 4.297 & 1277         &  75.7 &    1239 &       &           &               &               \\
\hline 
\end{tabular}
\begin{flushleft}
\noindent $^{\rm a}$The 4e-CCSD(T)+iTQ+OC calculation only 
for 9 internuclear distances ($R$) from 6.5 to 12~a.u.\ 
(and for 100~a.u.) which are presented in \tref{tabca2}.
\end{flushleft}
\end{table}

\clearpage

\end{document}